\documentclass[11pt]{article}
\usepackage{amssymb}
\usepackage{amsmath}
\usepackage{lscape}
\usepackage[dvips]{epsfig}
\def\nuc#1#2{\relax\ifmmode{}^{#1}{\protect\text{#2}}\else${}^{#1}$#2\fi}

\makeatletter
\newcommand{\l@vveden}[2]{\hbox to\textwidth{{\bf \quad #1 #2}}}

\makeatother

\begin{document}

{\bf \Large 
\noindent{
LANL Report LA-UR-00-3598, Los Alamos (2000)\\
}}

{\bf \Large 
\noindent{
Proc. SATIF5, July 18-21, 2000, Paris, France}}

\vspace*{1cm}

\begin{center}
{ \bf 
SPECTRAL MEASUREMENTS OF NEUTRONS FROM Pb, W, AND Na TARGETS IRRADIATED BY
0.8 AND 1.6 GEV PROTONS}

\vspace*{1.2cm}

{\bf  Yury E. Titarenko, Oleg V. Shvedov, Yury V. Trebukhovsky, \\ 
Vyacheslav F. Batyaev, Valery M. Zhivun, Dmitry V. Fishchenko, \\
Vladimir A. Korolyov, Gennady N. Smirnov }\\
{SSC Institute for Theoretical and Experimental Physics, Moscow, 117259, 
Russia}

{\bf Stepan G. Mashnik, Richard E. Prael, Arnold J. Sierk} \\
{Los Alamos National Laboratory,  Los Alamos, NM 87545, USA}

{\bf Hideshi Yasuda} \\
{Japan Atomic Energy Research Institute, Tokai, Ibaraki, 319-1195, Japan}

\end{center}

\vspace*{1.2cm}

\section*{Abstract}
\noindent

Preliminary results of neutron spectra measurements 
from
Pb, W, and Na targets irradiated by 0.8 and 1.6 GeV protons are presented. 
Measurements have been carried out using the proton beam extracted from the
ITEP synchrotron and the TOF technique. Neutron registration has been carried 
out using BICRON MAB-511 liquid scintillation counters. Spectra measured 
at angles of 30$^o$, 60$^o$, 90$^o$, 120$^o$, and 150$^o$ have been 
compared with results of their simulation using the LAHET code system
and the code CEM2k. 
The results are of interest both from data gathering viewpoint and as a 
benchmark of the up-to-the-date 
predictive powers of codes applied to design the hybrid Accelerator Driven 
Systems (ADS) using lead or lead-bismuth targets 
and sodium-cooled targets.

\section*{Foreword}
\noindent

Data on neutrons and charged particles generated from proton beam 
interactions with targets and structure materials are necessary when 
designing the present-day ADS facilities with proton beam 
energy of $\sim$1-2 GeV \cite{bib1, bib2}. Requirements to the data 
accuracy are rather strict because such data determine the 
external source term of the ADS. Besides, the neutron and proton data 
determine the calculation accuracy requirements of such principal ADS 
blanket parameters as the $k_{eff}$, safety control system efficiency, 
energy deposition of the fuel assembly, and the minor actinide 
transmutation rates. These data are also important in calculating 
radiation resistance of structure materials exposed to high-energy particles.

What is said above determines the necessity for further experimental 
investigations of particle generation cross sections and conducting 
more accurate measurements of these cross sections at energies 
of bombarding protons up to several GeV. Such results are important, 
first, as nuclear constants by themselves and, second, in verifying 
the computational codes used in practice to calculate the 
parameters of ADS facilities.

All known experiments in measuring double differential cross sections 
of neutrons generated as a result of interaction of protons of intermediate 
energies with thin and thick targets made 
of different materials are tabulated in Table \ref{tab1}.

An analysis of the data presented in these works shows that 
double differential cross sections of neutrons for lead measured at 
proton energy of 0.8 GeV at LANL, KEK, and SATURNE agree well with 
each other. They agree rather well with results of calculations performed 
with different codes. The agreement is worse for targets with small mass 
numbers where discrepancy may reach 100\%. With incident proton energy 
increased to several GeV, the discrepancy between 
experimental and calculation data increases too.

Additional measurements of neutron spectra 
and yields in the proton energy range up to 2 GeV for different materials 
are necessary 
to study  causes of mentioned discrepancies
and improve further the available models and codes. 
Such experiments for measuring neutron 
double differential cross sections from
Pb(p,xn), W(p,xn), and Na(p,xn) reactions in thick targets bombarded 
by protons with energies of 0.8 and 1.6 GeV were performed at 
the Institute of Theoretical and 
Experimental Physics (ITEP), Moscow. Measurements were made 
by the time-of-flight (TOF) techniques, neutron spectra were measured at 
angles of 30$^o$, 60$^o$, 90$^o$, 120$^o$, and 150$^o$ in the laboratory 
frame of reference. 

The data obtained were compared with results of calculations by the LAHET 
code system \cite{bib17} and the code CEM2k \cite{cem2k}.

\begin{table}
\caption{Neutron spectra experiments at proton energies above 100 MeV}

\vspace*{0.5cm}
\begin{tabular}{|c|p{3.2cm}|c|c|c|c|c|}\hline \label{tab1}
E$_{inc}$, MeV	&	Target nuclei &Neutron & Laboratory & Institute / & Refs.	\\ 
	&	 & energy, MeV &  angle, degrees &  Year	& 	\\ \hline
585	&	C, Al, Fe, Nb, In, Ta, Pb, U	&	0.9 -- E$_{max}$	&	30, 90, 150	&	PSI / 87	&	\cite{bib3}	\\ \hline
120, 160	&	Al, Zr, Pb	&	$\ge$ 30	&	0 -- 145	&	IUCF / 90	&	\cite{bib4}	\\ \hline
113	&	Be, C, O, Al, Fe, W, Pb, U	&	0.5 -- E$_{max}$	&	7.5 -- 150	&	LANL / 89	&	\cite{bib5}	\\ \hline
256	&	Be, C, O, Al, Fe, Pb, U	&	0.5 -- E$_{max}$	&	7.5 -- 150	&	LANL / 92	&	\cite{bib6}	\\ \hline
256, 800	&	Li, Al, Zr, Pb	&	20 -- E$_{max}$	&	7.5 -- 150	&	LANL / 93	&	\cite{bib7}	\\ \hline
318, 800	&	Al, Pb, U	&	5 -- E$_{max}$	&	7.5, 30	&	LANL / 86	&	\cite{bib8}	\\ \hline
597	&	Be, B, C, N, O, Al, Fe, Pb, U	&	0.5 -- E$_{max}$	&	30 -- 150	&	LANL / 93	&	\cite{bib9}	\\ \hline
800	&	Be, B, C, N, O, Al, Fe, Cd, W, Pb	&	0.3 -- E$_{max}$	&	30 -- 150	&	LANL / 92	&	\cite{bib10}	\\ \hline
800, 1500, 3000	&	C, Al, Fe, In, Pb	&	1 -- E$_{max}$	&	15 -- 150	&	KEK / 97	&	\cite{bib11}	\\ \hline
2200	&	Cu	&	3.3 - 200	&	60	&	KEK / 83	&	\cite{bib12}	\\ \hline
500, 1500	&	Pb	&	1 -- E$_{max}$	&	15 -- 150	&	KEK / 95	&	\cite{bib13}	\\ \hline
800, 200, 1600	&	C, Fe, Zr, Pb, Th	&	2 -- E$_{max}$	&	0 -- 160	&SATURNE/98	&	\cite{bib14}	\\ \hline
600 - 1600	&	Al, Cu, Zr, Pb	&	3 -- 200	&	30 -- 150	&	ITEP / 96	&	\cite{bib15}	\\ \hline
750, 1280, 2200	&	Cu, Pb, U	&	7.5 -- 70	&	119	&	ITEP / 83	&	\cite{bib16}	\\ \hline
  \end{tabular} 
\end{table}

\section*{Description of the experiment}
\noindent

\begin{figure}[t] 
\centerline{\epsfxsize 11cm \epsffile{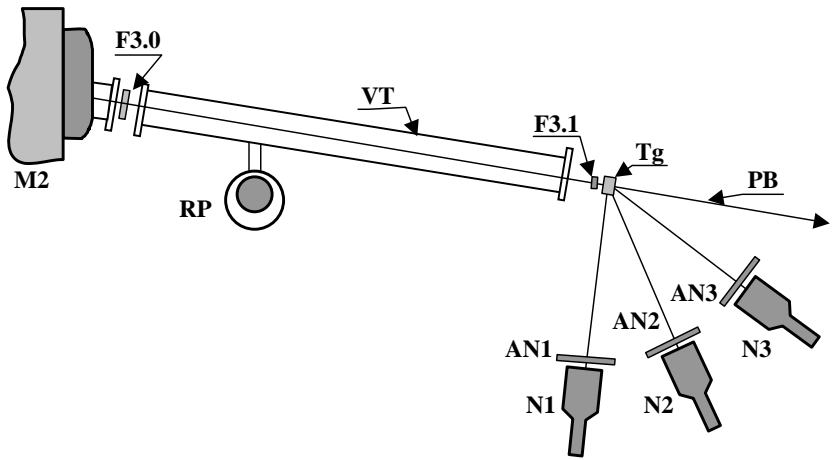}} 
\caption{The experimental facility layout.}\label{fig1} \end{figure}

The experiment has been carried out using the time-of-flight (TOF)
technique, the TOF spectrometers were located in the
512nd beam of the ITEP proton synchrotron with a
maximum energy of 10 GeV. Detectors were located 
at a distance of 2.5 m from the floor and more than 5 m from the ceiling 
and walls. The beam intensity 
was of approximately 10$^5$ protons per pulse. The beam was focused at 
the center of the investigated 
targets, its profile was close to an ellipse with axes of 
2 cm $\times$ 2.5 cm. The distance between the 
target and neutron detectors changes from 1.5 m to 3 m and is not evacuated. 
The target materials and sizes are listed in Table \ref{tab2}. 
The contents of impurities in tungsten and sodium were less than 0.2\% and
0.02\%, respectively. Sodium was placed in a cylindrical steel container 
with 0.4-mm thick walls.
The experimental facility layout is shown in Fig. \ref{fig1}, where 
PB is the proton beam, M2 is the bending magnet, Tg  is the target 
under investigation, F3.0 and F3.1 are plastic 
scintillators. 

A 12-m distance was selected to minimize the effect of the great mass 
of large magnet M2 on the measurement results. The targets under 
investigation were located in the second 
focus 
of the beam at 80 m from the accelerator internal target.

The particles leaving the target are recorded by three detector 
assemblies (N1, N2, N3). Each of the assemblies consists of a 
1 cm x 19 cm x 19 cm plastic scintillator (AN1, AN2, AN3) placed 
in the immediate proximity 
to, and ahead, a BICRON MAB-511 $\varnothing$ 12.7 cm $\times$ 15.2 cm 
liquid neutron detector. 
There was no protection of the neutron detectors. Scintillators were 
turn on for coincidence with neutron detectors in charged particles 
spectra measurements and for anti-coincidence, 
in neutron spectra measurements.

\begin{figure} 
\begin{center}
\includegraphics[angle=-90, width=16cm]{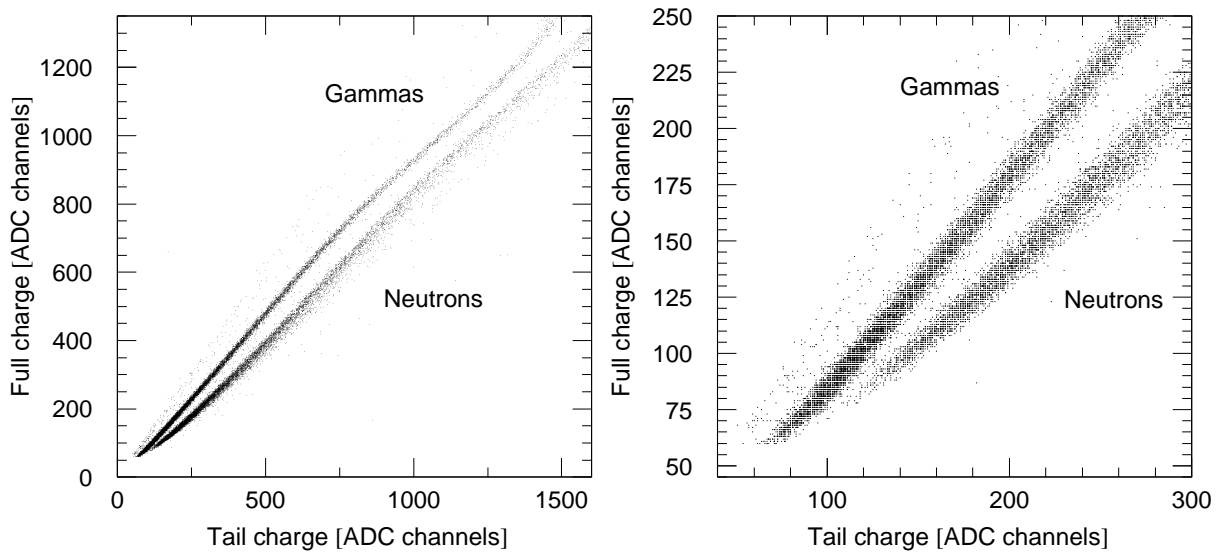}
\caption{Separation of neutrons and gammas with the
amplitude-amplitude analysis.}
\label{fig2} \end{center} \end{figure}

\begin{figure} 
\begin{center}
\includegraphics[angle=-90, width=12cm ]{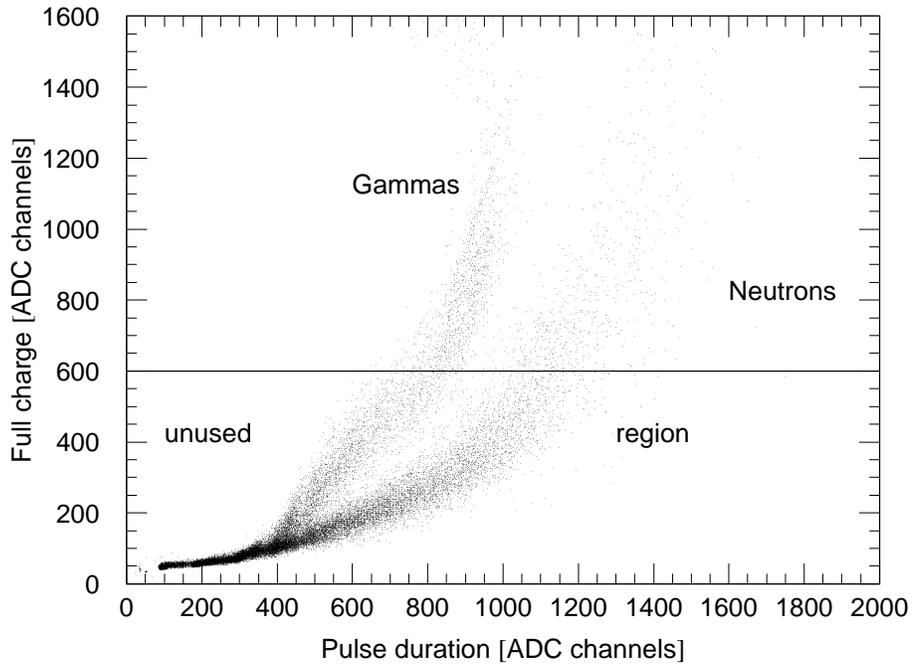}
\caption{Separation of neutrons and gammas with the
amplitude-time analysis.}\label{fig3}  \end{center}  \end{figure}

Separation of neutrons and gammas have been performed with an
amplitude-amplitude analysis of the registered particle pulse 
(A(full charge) -- A(tail charge)) within the recoil proton 
energy range of 2.5 -- $\sim$10 MeV and an amplitude-time 
(A(full charge) -- T(pulse duration)) analysis within the recoil 
proton energy range of $\sim$ 10 -- 300 MeV. The first method 
provides reliable separation of the small 
amplitude pulses. This is shown in Fig. \ref{fig2}. The second method 
separates large amplitude pulses 
(Fig. \ref{fig3}), where the quality of the amplitude-amplitude 
separation is lost.

\begin{table}
\begin{center}
\caption{The target materials and sizes}

\vspace*{0.5cm}
\begin{tabular}{|c|c|c|}\hline \label{tab2} 
Target material & Proton energy, GeV & Target size, cm \\ \hline
Pb & 0.8 &  $\varnothing$ 6.0 $\times$ 2.0 \\ \hline
Pb & 1.6 & 15$\times$15$\times$20 \\ \hline
W & 0.8, 1.6 & $\varnothing$ 5.0 $\times$ 3.0 \\ \hline
Na & 0.8, 1.6 & $\varnothing$ 6.0 $\times$ 20 \\ \hline
 \end{tabular} 
 \end{center}
\end{table}

Fig. \ref{fig3} demonstrates the branch behavior of the 
amplitude-time separation 
technique.  
From Fig. \ref{fig3}, one can see that with increasing the pulse 
amplitude the branches that corresponds to neutrons and gammas 
diverge and the quality of separation increases accordingly.

Thus, an acceptable quality of separation was achieved in the range 
of small pulse amplitudes by appropriate matching the parameters, 
and in the range of large pulse amplitudes, by using the 
amplitude-time separation technique.

\begin{figure} 
\begin{center}
\includegraphics[angle=-90, width=10cm]{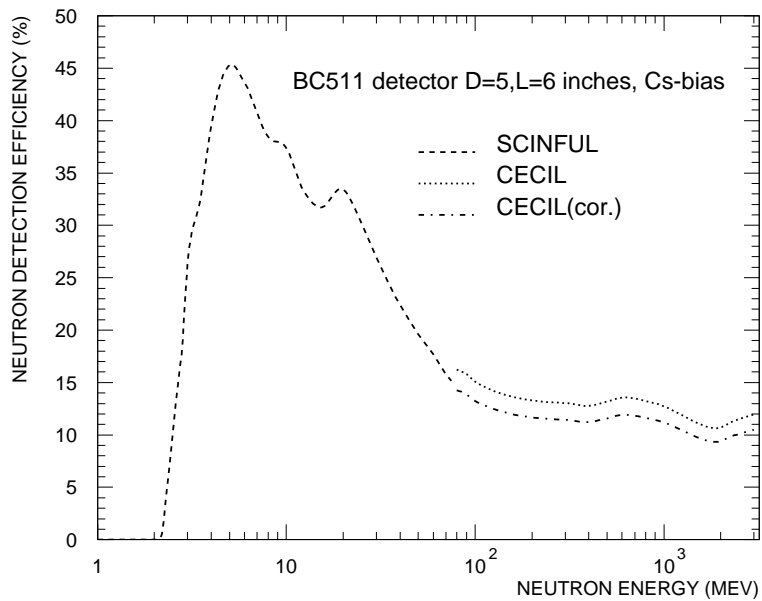}
\caption{Efficiency of the BC511 detector used for neutron registration 
(detector size d5$'' \times $ L6$''$). 
Calculation has been performed using the SCINFUL and CECIL codes at 
threshold corresponding to \nuc{137}{Cs}.}\label{fig4}  
\end{center} 
\end{figure}

The neutron counter efficiency was calculated using the SCINFUL \cite{bib18} 
and  CECIL~ \cite{bib19} codes. 
Because the SCINFUL code application is limited to 80 MeV and the 
CECIL code gives reliable results 
up to energies of several hundred of MeV, the results of calculation 
with the SCINFUL code were used for energies below 80 MeV and the results 
of calculation with the CECIL code were used 
for higher energies. The results of calculation using the CECIL code 
at 80 MeV and above
were renormalyzed for 
matching with the results of the SCINFUL code at 80 MeV 
(see Fig. \ref{fig4}). The error in determining 
the efficiency is estimated to be equal to 10\% at energies below 80 MeV 
and 15\% at higher energies.

\section*{Simulation of neutron spectra}
\noindent

Because targets used in the present experiment can not be regarded as thin,
simulations of neutron spectra by
LAHET have included not only neutron generation from
the proton-nucleus interactions, 
but also multiple 
scattering of primary protons together with the low energy (below 20 MeV) 
neutron transport by the
HMCNP code. In the cases of lead and tungsten,
elastic scattering of neutrons with energy above 20 
MeV was taken into consideration as well.

\section*{Results}
\noindent

The measured neutron spectra from lead, tungsten, and sodium for 
proton energies of 0.8 and 1.6 GeV at angles of 30$^o$, 60$^o$, 90$^o$, 
120$^o$, and 150$^o$ are shown in Figs. 5-7. The experimental data 
from other works (\cite{bib10} and \cite{bib11}, for Pb at 0.8 GeV; 
\cite{bib14}, for Pb at 1.6 GeV; 
\cite{bib10}, for W 0.8 GeV) and calculations by
 LAHET are shown in the figures as well.

Comparison of experimental and calculation results shows
a satisfactory agreement for the
heavy nuclei targets, W and Pb (Fig. \ref{fig8}),
at both proton energies. 
Exceptions may be seen for neutrons with energy above 100 MeV at angles  
60$^o$, 90$^o$, and 120$^o$ for $T_p =1.6$ GeV
and for energetic neutrons at 
90$^o$, 120$^o$, and 150$^o$ for $T_p =0.8$ GeV. The agreement
of calculated results with the data is 
worsen with transfer to sodium. Traditionally, this is explained 
by problems for the most of 
theoretical models to describe high-energy hadron interactions 
with nuclei of low masses.

As an example, for the lightest element measured, Na, where the 
thickness of target should be of the least importance for the measured 
neutrons, we show also calculations with the last version of the 
Improved Cascade-Exciton Model code, CEM2k \cite{cem2k}, simulating 
pure proton-nucleus reactions, without taking into account any 
internuclear interactions (Fig. \ref{fig7}). One can see that for 
neutron energies above several MeV, where the thickness of target no 
longer affects significantly the measured spectra, 
CEM2k agrees with the data quite well, though some discrepancies in 
the very tails of the spectra still remain to be understood. 
Calculations with LAHET (both ISABEL and Bertini options) take into account 
the thickness of targets, therefore agree somewhere better than CEM2k 
with this data. Nevertheless, some disagreements between LAHET results
and the data at the high-energy tails of most spectra and 
around $\sim 20$ MeV at forward angles for 
Na have yet to be understood. At a glance, it looks like we got with 
both LAHET and CEM2k too many preequilibrium 
neutrons at forward angles and too few high-energy neutrons at 
backward angles; the last could be an indication that the local Fermi 
distribution for intranuclear nucleons used by all models may be a 
too rough approximation. But these points need a further, more 
detailed investigation.

\begin{figure} 
\begin{center}
\includegraphics[angle=-90, width=16cm ]{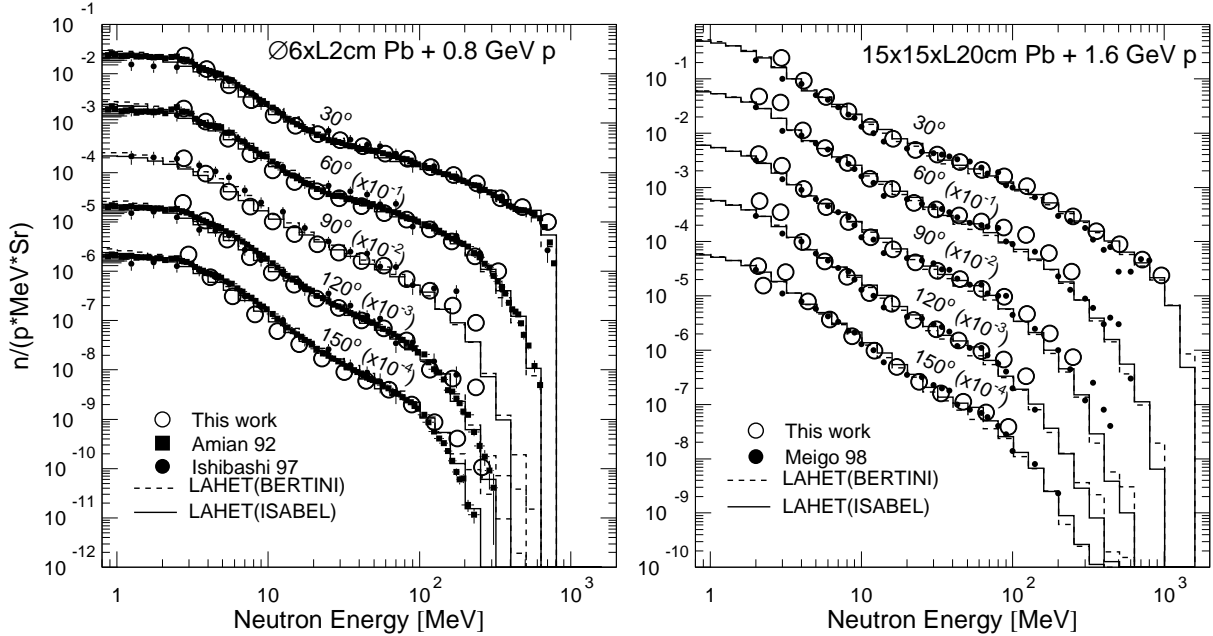}
\caption{Double differential neutron spectra from $^{nat}$Pb 
at proton energies of 0.8 and 1.6 GeV measured in the present work ($\circ$),
in previous works ([10] and [11], for 0.8 GeV and [14], for 1.6 GeV),
together with the results of calculation by LAHET.}
\label{fig5}  
\end{center} \end{figure}
 
\begin{figure} 
\begin{center}
\includegraphics[angle=-90, width=16cm]{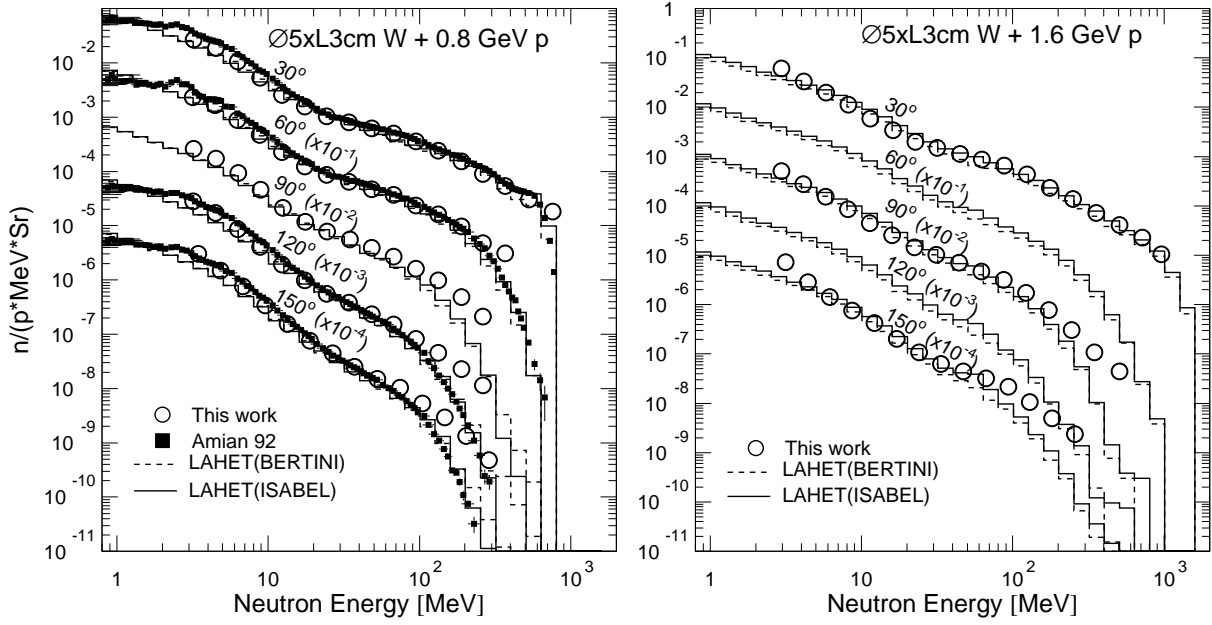}
\caption{Double differential neutron spectra from $^{nat}$W 
at proton energies of 0.8 and 1.6 GeV measured 
in the present work ($\circ$), in previous a work ([10], for 0.8 GeV),
together with calculations by
LAHET.}
\label{fig6}  
\end{center}  
\end{figure}
 
\begin{figure} 
\begin{center}
\includegraphics[angle=-90, width=16cm]{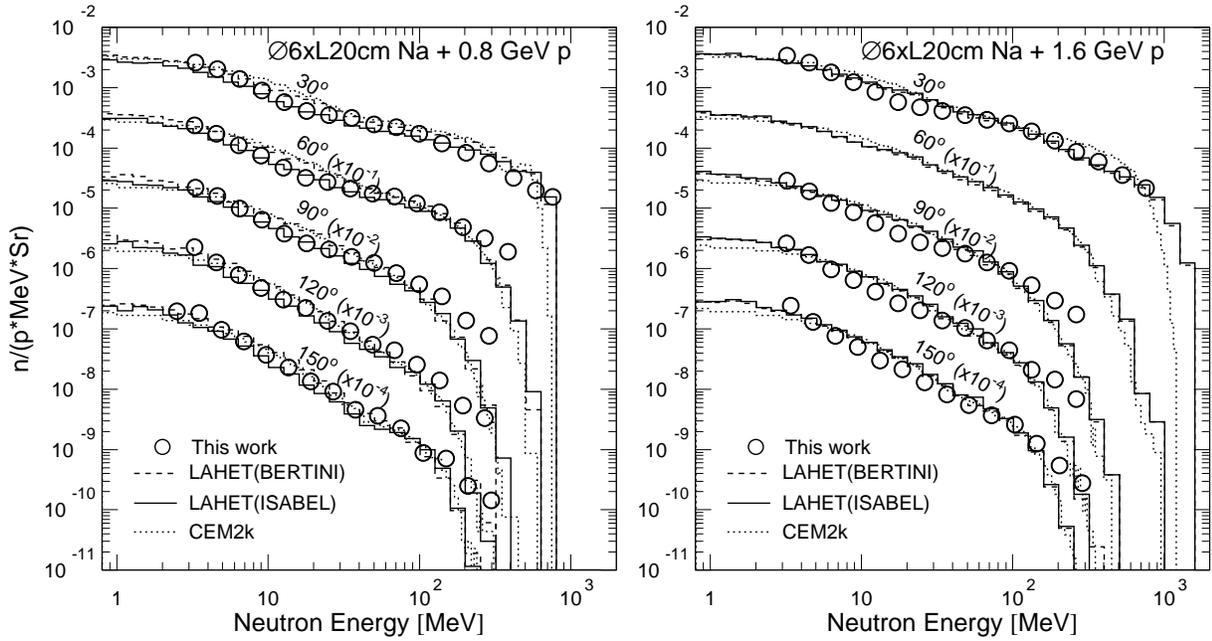}
\caption{Double differential neutron spectra from Na at proton
energies of 0.8 and 1.6 GeV measured 
in the present work ($\circ$) together with the result of calculations
by LAHET and CEM2k codes.}
\label{fig7}  
\end{center} 
\end{figure}
 
\begin{figure} 
\begin{center}
\includegraphics[angle=-90, width=12.5cm]{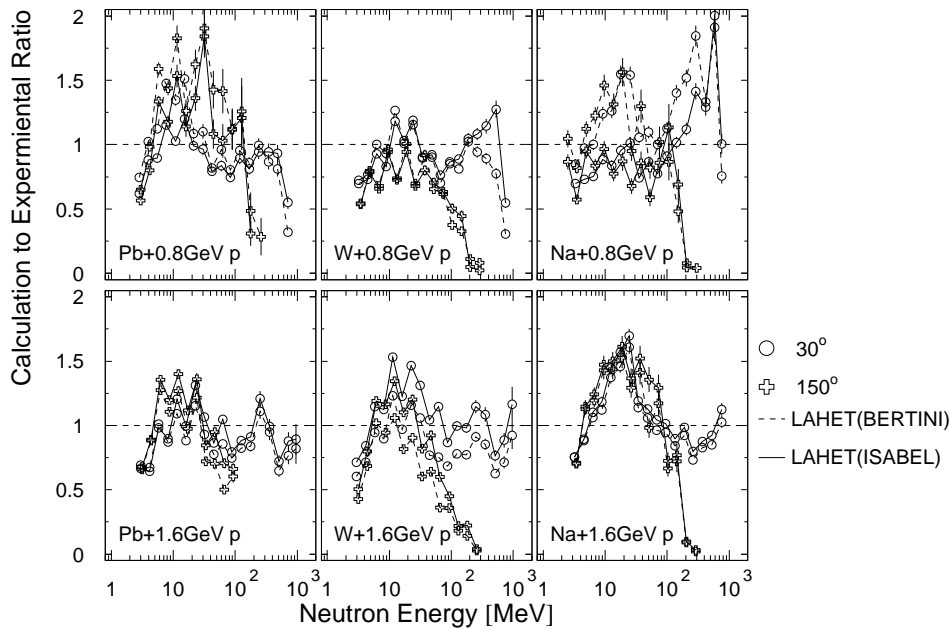}
\caption{Calculation-to-Experimental data ratios
of neutron spectra at 30$^o$ and 150$^o$ from
Pb, W, and Na, as indicated.}
\label{fig8} 
\end{center} 
\end{figure}

\section*{Acknowledgement}
\noindent

The authors are indebted to Dr. S. Meigo (JAERI) for his assistance 
in calculating the detection efficiency, to Dr. O. A. Shcherbakov 
(PINP, Gatchina) for helpful discussions of the measurement techniques, 
to Dr. V. L. Romadanov (MEPhI, Moscow) for his 14 MeV neutron source
used in detector calibration, to Drs. V. N. Kostromin and 
I. A. Vorontsov (ITEP, Moscow) for their assistance in carrying out 
the experiments, and to Dr. F. E. Chukreev (KIAE, Moscow) for discussions 
of results.

The work has been carried out under the ISTC Project\#1145 and was
supported by the JAERI (Japan) and, partially, by the 
U. S. Department of Energy.

\end{document}